\documentclass[apj]{emulateapj}
\usepackage[usenames,dvipsnames]{xcolor}
\definecolor{blue}{rgb}{0,0,1}
\usepackage[hyperfootnotes=false]{hyperref}
\usepackage{amsmath,amssymb,amsthm}
\hypersetup{
  colorlinks,
  citecolor=blue,
  linkcolor=blue,
  urlcolor=Green}

\slugcomment{}

\shortauthors{Zhang, Lin, Burkert \& Oser}

\begin{document}

\title{Galacto-forensic of LMC's orbital history as a probe for
the dark matter potential in the outskirt of the Galaxy}
\submitted{Draft version \today}

\author{Xiaojia Zhang\altaffilmark{1,2}$^*$, Douglas N. C. Lin\altaffilmark{2,3},
Andreas Burkert\altaffilmark{4,5,6}, Ludwig Oser,\altaffilmark{7}}
\altaffiltext{*}{E-mail: \href{mailto:xiaojia.f.zhang@gmail.com}{xiaojia.f.zhang@gmail.com}}
\altaffiltext{1}{Department of Astronomy, Peking University, Beijing 100871, China}
\altaffiltext{2}{Kavli Institute for Astronomy and Astrophysics, Peking University, Beijing 100871, China}
\altaffiltext{3}{Department of Astronomy and Astrophysics, University of California, Santa Cruz, CA 95064, U.S.A.}
\altaffiltext{4}{University Observatory Munich, Scheinerstrasse 1, D-81679 Munich, Germany}
\altaffiltext{5}{Max-Planck-Institute for Extraterrestrial Physics, Giessenbachstrasse 1, 85758 Garching, Germany}
\altaffiltext{6}{Max-Planck Fellow}
\altaffiltext{7}{Max-Planck-Institute for Astrophysics, Karl-Schwarzschild-Str. 1, D-85741 Garching. Germany}

\begin{abstract}
The 3D observed velocities of the Large and Small Magellanic Clouds
(LMC and SMC) provide an opportunity to probe the Galactic potential
in the outskirt of the Galactic halo. Based on a canonical NFW model
of the Galactic potential, Besla et al.(2007) reconstructed LMC and SMC's
orbits and suggested that they are currently on their first
perigalacticon passage about the Galaxy.  Motivated by several recent
revisions of the Sun's motion around the Galactic center, we
re-examine the LMC's orbital history and show that it depends
sensitively on the dark-matter's mass distribution beyond its
present Galactic distance. We utilize results of numerical
simulations to consider a range of possible structural and
evolutionary models for the Galactic potentials.  We find that
within the theoretical and observational uncertainties, it
is possible for the LMC to have had multiple
perigalacticon passages on the Hubble time scale, especially
if the Galactic circular velocity at the location of the Sun
is greater than $\sim 228$km s$^{-1}$.  Based
on these models, a more accurate determination of the LMC's
motion may be used to determine the dark matter distribution
in the outskirt of the Galactic halo.
\end{abstract}


\section{Introduction}
The Large and Small Magellanic Clouds (LMC and SMC) are the two most
prominent satellite galaxies in the Local Group.  They are accompanied
by the Magellanic Stream which extends along a great circle over 100
degrees across the sky. One end of the Stream joins, both in position
and radial velocity, onto the bridge between the LMC and SMC
(Meatheringham et al 1988). A continuous velocity gradient
leads to a large infall speed at the other end of the Stream
(Mathewson, Cleary, \& Murray 1974).

The kinematic properties of the LMC, SMC, and the Magellanic Stream
have motivated many investigations on their origin. One class of
models are based on the assumption that the Stream is the tidal
debris of the LMC and SMC during their past perigalacticon passages
(Toomre 1972, Mirabel \& Turner 1973, Fujimoto \& Sofue 1976,
Lynden-Bell \& Lin 1977, Murai \& Fujimoto 1980, Gardiner et al 1994,
Putman et al 1999). Numerical simulations with idealized Galactic
potentials have led to the prediction for the LMC's proper motion
to be 0.02$''$ per century in a direction such that it leads the
Magellanic Stream (Lin \& Lynden-Bell 1982). A subsequent
ground-based observational confirmation gave a slightly smaller value
(Jones et al 1994, Lin et al 1995). Nevertheless, the tidal scenario
has been challenged by the lack of excess halo stars in the direction of
the Stream (Majewski et al. 2003) since in such a model, stars and gas
should have been equally stripped from the Magellanic Clouds.
Some of these issues may be resolved with an alternative scenario that
the gas in the Magellanic Stream was removed from the LMC by a ram pressure
as it ploughed through the hot residual gas in the Galactic halo
(Moore \& Davis 1994). The density of the halo gas must be sufficiently
high to remove atomic gas from the LMC. It also needs to be sufficiently
tenuous to avoid any drag on the Stream and reduction of its infall speed
below its observed values (Mastropietro et al 2005).

One possible test to distinguish between these scenarios is
to reconstruct the LMC's orbital history with a set of accurate
observational data. The proper-motion measurements by Kallivayalil
et al.(2006a, 2006b) made with the Hubble Space Telescope (HST)
provided a set of accurate 3D Galactocentric velocities for
both LMC and SMC.  After correcting for the motion of the Sun
(they adopted a Galactic circular velocity, at the Sun's location,
of $V_c=220$km s$^{-1}$),
they obtained a transverse velocity of the LMC to be
$\sim367$km s$^{-1}$ which is much larger than the inferred
Galactic circular velocity at its present-day location and its
radial velocity which has a positive, modest value of
$\sim89$km s$^{-1}$.

The newly obtained 3D velocity data confirm that the LMC has
just passed its perigalacticon. They are also useful for the
reconstruction of the LMC's orbital history especially at the
time of its previous perigalactic passage. But, such a
determination depends on the prescription for the Galactic
potential. For example, Lin \& Lynden-Bell (1982) adopted an
idealized model which is almost certain over simplified.
Although, the most recently measured proper motion is in agreement with
their prediction, to within a second decimal place, LMC's
orbit needs to be re-examined with a more appropriate Galactic
potential. Using these velocities and a $\Lambda$CDM-motivated
MW model with the virial radius ($R_{vir}=258kpc$) and mass
($M_{vir}=10^{12}M_\odot$), Besla et al.(2007) reconstructed
the orbital history of the Clouds and suggested that they are
on their first passage about the MW.

The first-passage conclusion obtained by Besla et al.(2007), if
verified, would invalidate the tidal origin of the Magellanic
Stream. In a follow-up investigation, Besla et al. (2010) suggested
that gas in the Magellanic Stream was torn from the SMC by the
tide of LMC before the two Clouds impinged to the apogalacticon.
To match the observed column density along the Stream, they
adopted a gas to star (and dark matter) mass ratio for the SMC
to be an order of magnitude larger than those they adopted for the
LMC.  In order to distinguish these alternative scenarios, it is
desirable to check the robustness of their results.

We begin with a re-analysis of the kinematic
data. At the Clouds' present-day position in the sky, a large
fraction of their observed line of sight and proper motion
speeds are due to the Sun's motion around the Galactic center.
The distance of the Sun from the Galactic center and the
circular velocity of the local standard of rest have been
revised recently (Reid et al. 2009a) to be $R_0=8.4\pm0.6kpc$
and $V_c=254\pm16$km s$^{-1}$ respectively. These latest
determinations are based on the reference frame set by the
maser sources near the Galactic center. In \S2, we show that
this new value of $V_c$ modifies the correction of the solar motion
and LMC's 3D speed. It also significantly revises the reconstruction
of the Magellanic Clouds' orbital history (see a similar conclusion
by Shattow \& Loeb 2009).

But, there are other recent
analysis which lead to a range of values for
$V_c$'s.  Based on 18 precisely measured Galactic masers, Bovy
et al (2009) deduced a lower estimation for the circular velocity at
the Sun's location to be $V_c=236\pm11$ km s$^{-1}$. In another APOGEE
analysis, Bovy et al (2012) concluded $V_c=218\pm$ 6 km s$^{-1}$ and
claimed that $V_c < 235$ km s$^{-1}$ at $>99 \%$ confidence level.  In
light of this range of $V_c$, we will discuss, in \S4.2, the
dependence of the LMC's orbital history on $V_c$ within an error bar.

A high value (254 km s$^{-1}$) of $V_c$ also indicates that the
rotation curve
of the Milky Way Galaxy is similar to that of the Andromeda Galaxy,
suggesting they may have comparably massive dark matter halos.
Their combined masses can be estimated from the timing argument
(Kahn \& Woltjer, 1959) because these two most prominent galaxies
in the Local Group are currently approaching each other and they
are presumably bound to each other. Based on a recent measurement
of M31's proper motion, van der Marel et al (2012) estimate a
total mass of the Local Group to be $3 \sim 4 \times 10^{12}M_{\odot}$.
If their kinematic properties are very similar to each other,
the mass of either the Milky Way or M31 would be $1.5\sim2
\times 10^{12}M_{\odot}$.  This mass is mostly in the form of a
dark matter halo.  In order to verify this possibility, we
re-analyze the space motion of the LMC based on this new
information. We briefly recapitulate the equation of motion in \S2.

In \S3, we illustrate that the LMC's orbital history also depends
sensitively on the mass distribution in the outer Galactic halo,
over and above the uncertainties introduced by the magnitude of
$V_c$.  As an illustration, we prescribe
the Galactic potential as $\Phi\propto{r^{-\lambda}}$ ($0<\lambda<1$)
and show that if $\lambda\sim0.1$, the LMC would be currently on
its second passage about the Milky Way.  It is useful to adopt
a more realistic Galactic potential and determine the
history of the LMC's orbit.

There are several attempts to reconstruct the Galactic
potential based on the velocity information of satellite galaxies
(Peebles 2001, 2010). However these determinations are highly
uncertain.  We follow the approach of Besla et al (2007)
by utilizing an idealized potential based on the $\Lambda$CDM
simulations.  These simulations produce to well determined profiles
for the density distribution in the inner regions of galaxies
(Navarro et al. 1997, hereafter NFW).  However, our illustrative
model indicates that LMC's past orbit sensitively depends on the
potential in the outer regions of the galaxy. Numerical simulations
show that there are considerable uncertainties and dispersion in
the mass distribution for outer regions of galaxies.

In order to explore the possible range of the LMC's orbital history,
we select, in the \S4, several sample dark-matter
halos of galaxies from numerical CDM simulations.  These models
have slight variations in the slope of the dark matter density
distribution. We also take into consideration that with the
revised circular velocity of the Sun's motion, the mass of the
Galaxy may need to be upgraded to be comparable to that of M31.
With the simulated profile of the Galactic potential, we show
that it is possible for the LMC to have had several encounters
with the Galaxy.  Since the mass ratio between the LMC and SMC is
$10:1$, we neglect SMC's contribution. In \S5, we summarize our
results and discuss their implications.

\section{Orbital parameters}

\subsection{LMC's spacial velocity}

In their investigation, Kallivayalil et al. (2006a) used
$V_{c, sun}=220$km s$^{-1}$ as the circular velocity of the Sun.  (This
value is similar to that derived by Bovy et al. 2012).  They
followed the method outlined in van der Marel et al. (2002),
and obtain the LMC's total, radial, and azimuthal velocities as:
$v_{LMC}=378\pm18$km s$^{-1}$, $v_{LMC,rad}=
89\pm4$km s$^{-1}$, and $v_{LMC,azi}=367\pm18$km s$^{-1}$
respectively.  Here we adopt the same procedure to correct
for the solar motion, including the Sun's peculiar motion relative
to the local standard of rest determined by Dehnen \& Binney (1998)
as $(U_\odot, V_\odot, W_\odot)=(10.0\pm0.4, 5.2\pm0.6, 7.2\pm0.4)km/s$,
and apply the recently revised $V_{c,sun}\sim250$km s$^{-1}$ (Reid
et al. 2009a) to compute the LMC's total, radial, and azimuthal
velocities. We adopt this high value for $V_{c, sun}$ to illustrate
that it can significantly modified the deduced orbital history of
the Magellanic Clouds (also see, Shattow \& Loeb 2009).  For this revised
value of the solar motion we find values of $v_{LMC}\sim356$km s$^{-1}$,
$v_{LMC,rad}\sim63$km s$^{-1}$, and
$v_{LMC,azi}\sim351$km s$^{-1}$ respectively.

\subsection{LMC's equation of motion}
All of the new kinematic quantities (derived with the newly
revised $V_{c, sun}$ are one sigma smaller than their previously
determined values and they are expected to imply a more bound
LMC's orbit around the Galaxy.  In order to verify this
expectation, we compute the LMC's orbital history.

Under the influence of the Galactic potential and dynamical friction,
LMC's equation of motion is

\begin{equation}
\ddot{\textbf{r}}=\frac{\partial}{{\partial}\textbf{r}}\Phi(r)
+\frac{\textbf{F}_{DM}}{M_{LMC}}.
\end{equation}
We adopt the idealized Chandrasekhar(1943) formula for dynamical
friction such that
\begin{equation}
\textbf{F}_{DM} = - \frac{4{\pi}{ln\Lambda}G^2M_{LMC}^2
\rho(r)}{V_{LMC}^3}[erf(X)-\frac{2X}{\sqrt{\pi}}e^{-X^2}]\textbf{V}_{LMC}
\end{equation}
where $X{\equiv}V_{LMC}/\sqrt{2}\sigma$, $\sigma$ is the velocity
dispersion of the DM halo and $\rho(r)$ is the density of the halo at r.
For the present application, we neglect the effect of an active halo
and mass loss from the LMC (Fellhauer \& Lin 2007). The mass of the
LMC is about $2 \times 10^{10}M_{\odot}$. In most previous applications,
the Coulomb logarithm in the above formula is assumed to be a constant
$\Lambda=b_{max}/b_{min}$ (Binney and Tremaine 1987). Here we use
a prescription introduced by Hashimoto et al.(2003) in which $b_{max}$
is replaced by the position of the satellite and $b_{min}=4.8kpc$.
After the specification of the
Galactic potential, equation (1) can be solved numerically with
a symplectic leapfrog integration scheme(Springel et al. 2001).
We verify the numerical accuracy with
a range of time steps. For typical models, several hundred steps
per orbit are more than sufficient for numerical convergence.

\section{Dependence on potential model of the Milky Way}
We first show the sensitive dependence of the LMC's orbital period
on the Galactic potential. For illustration purpose, we adopt a
spherically symmetric, static model with a power law potential
profile for the Galactic potential.

\subsection{A simple power law profile}
At LMC's Galactocentric distance, the local circular velocity is
much larger than its radial velocity and smaller than its transverse
velocity, and the radial velocity is positive. These kinematic
properties suggest that the LMC has
just passed its perigalacticon. Since the LMC is close to its
perigalacticon, its orbital history is mainly related to the structure
of the Galaxy outside the perigalacticon distance of LMC. Therefore,
we mainly focus on the outer part of the Galaxy in order to estimate how
the orbital period of the LMC depends on the model parameters.

We assume an idealized power-law potential
$\Phi\propto{r^{-\lambda}}$($0<\lambda<1$).  The gravity felt by the LMC
at a distance $r$ is
\begin{equation}
F(r)=\frac{GM(r_p)M_{LMC}}{r_p^2} \left(\frac{r}{r_p}
\right)^{-\lambda-1}
\end{equation}
where $r_p$ is the perigalacticon distance of the LMC in Galactocentric
Coordinates and $M(r_p)$ is the total mass of the Galaxy inside $r_p$.
Integrate this force to get the potential of the Galaxy beyond $r_p$:
\begin{equation}
\Phi(r)=\Phi(r_p)\left(\frac{r}{r_p}\right)^{-\lambda}=\frac{1}{\lambda}
\frac{GM(r_p)}{r_p}\left(\frac{r}{r_p}\right)^{-\lambda}
\end{equation}

Under the conservation of energy and angular momentum,
\begin{equation}
\frac{1}{2}\left[1-\left(\frac{1-e}{1+e}\right)^2\right]
\left[1-\left(\frac{1-e}{1+e}\right)^\lambda\right]^{-1}
=\frac{\Phi(r_p)}{v_t^2}
\end{equation}
where the orbital eccentricity is defined such that
$r_p/r_a = (1-e)/(1+e)$, $v_t$ is the tangential velocity
of the LMC at $r_p$. Substitute $\Phi(r_p)$ and define:
\begin{eqnarray}
A (e,\lambda) &\equiv& \frac{\lambda}{2}
\left[1-\left(\frac{1-e}{1+e}\right)^2\right]
\left[1-\left(\frac{1-e}{1+e}\right)^\lambda\right]^{-1}  \notag \\
&=&\frac{GM(r_p)}{v_t^2r_p}=\frac{v_c(r_p)^2}{v_t^2}.
\label{eq:fofe}
\end{eqnarray}
In the above equation, the magnitude of $A$ is determined by
the LMC's azimuthal velocity and distance at perigalacticon.
For a given $\lambda$, each value of $A$ implies a set of values
for $e$, $r_a$, and orbital period (see fig 1). For a given value
of $\lambda$, $r_a/r_p$ would decrease rapidly with increase of A
if the LMC's eccentricity is high. It would also decrease
rapidly with decreasing value of $\lambda$ if the value of $A$
is close to the minimum value required for a bound orbit.
These tendencies imply that, in the limitation
of high eccentricity, a small modification in the value
of e (due to fractional changes in the values of $A$ or
$\lambda$) would lead to very different orbital periods.

The derivation of equation (6) is based on the assumed
conservation of energy and angular momentum. Had we taken into
account the effect of dynamical friction, the loss of energy
and angular momentum in time would imply a larger apogalacticon
distance and a longer orbital period in the past, albeit the
magnitude of its contribution is modest. The uncertainties of
the LMC's transverse velocity $v_t$ and mass distribution
of the Galaxy are contained in the values of $A$ and $\lambda$.
In the previous section, we have already discussed the
potential reduction in the value of $v_t$ due to an upward
revision in the value of $V_{c, sun}$.  For a given value
of $\lambda$,  the corresponding increase in $A$ (eq. 6)
would reduce the values of $r_a/r_p$ (Fig. 1) and
LMC's Galactic orbital period.

\begin{figure}
  \centering
  \includegraphics[width=0.9\linewidth,clip=true]{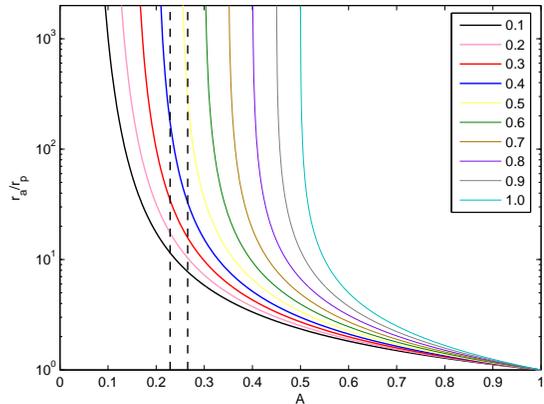}
\caption{The dependence of ratio $r_a/r_p$ on $A$ with different
values of $\lambda$. From below to above the 10 lines corresponds
to $\lambda$ from $0.1$ to $1.0$ respectively. The two vertical dashed lines represent the value span of A from 0.229 to 0.265 corresponding to $V_t=378km/s$ to $351km/s$ for LMC.}
  \label{fig:fig1}
\end{figure}

Adopting the most recently revised circular velocity of the
local standard of rest and LMC's location ($r_p\sim50kpc$)
as $v_t$ and $r_p$, equation (1) can be solved for different
values of $\lambda$.  In order to illustrate this point,
we choose $\lambda=0.1$ and $M(r_p)=3.8 \times 10^{11}M_\odot$
(model $A_1$ of Klypin et al. 2002)
to examine the orbital history. This value of $M(r_p)$ is
identical to that used by Besla et al (2007). With these parameters the corresponding value of A for LMC is $0.229$ and $0.65$ under the tangential velocity as $V_t=378km/s$ and $V_t=351km/s$ respectively(Fig. 1).
 We choose a small
value for $\lambda$ because if the rotation curve remains
relatively flat in the outside part of the Galaxy, $\lambda$
would be very close to zero.

In order to determine the magnitude of dynamical friction
from equation (2), we adopt the density distribution of the
Galactic dark matter halo from equation (4). We choose
the largest value ($1$) for $X$.  This approximation slightly
over-estimates the contributions from the dynamical friction.
With this form of the potential, we obtain more than
one past perigalacticon passage by the LMC for both the
previous (which yields $v_t =378$km s$^{-1}$) and the
latest solar circular velocities (which yields
$356$km s$^{-1}$) (Figure 2).  The smaller magnitude of
the newly determined azimuthal speed decreases the LMC's
orbital period significantly as expected.

\begin{figure}
  \centering
  \includegraphics[width=0.9\linewidth,clip=true]{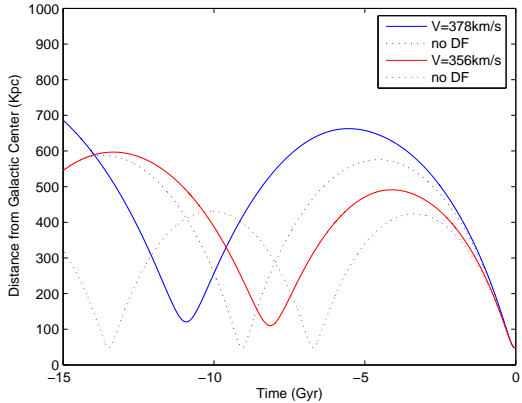}
\caption{The dependence of orbital history on the LMC's velocity
with $\lambda=0.1$. The dotted lines are the results without
dynamical friction.  Both velocities would lead to more than
one passages during Hubble time even with under the influence
of a strong dynamical friction effect.}
  \label{fig:fig2}
\end{figure}

\section{NFW model and simulation data}

\subsection{Models of Galactic potential}
In order to construct a more realistic Galactic potential, we
separate the baryonic and dark matter contribution to the
Galactic potential such that
\begin{equation}
M(r)=M_b(r)+M_{dm}(r)
\end{equation}
We assume that the dark matter density profile is described by
a NFW model (Navarro et al. 1997):
\begin{equation}
\rho_{halo}(r)=\frac{\rho_s}{x(1+x)^2}, \ \ \ \ \ \ x=\frac{r}{r_s},
\label{eq:nfwden}
\end{equation}
\begin{equation}
M_{halo}(r)=4\pi\rho_s r_s^3f(x)=M_{vir}f(x)/f(C)
\end{equation}
\begin{equation}
f(x)=ln(1+x)-\frac{x}{1+x}
\end{equation}
\begin{equation}
C=r_{vir}/r_s
\end{equation}
\begin{equation}
M_{vir}=\frac{4\pi}{3}\rho_{cr}\Omega_M\delta_{th}r_{vir}^3
\end{equation}
In the above equations, C is the halo concentration, $M_{vir}$ and
$r_{vir}$ are the halo virial mass and radius. The $\rho_{cr}$ is
the critical density of the universe and $\delta_{th}$ is the
overdensity of a collapsed object according to the spherical top-hat
model(Gunn \& Gott 1972). The $r_{vir}$ is defined as the radius
inside which the average density equals to the virial density
($\rho_{vir}=\delta_{th}\rho_M=\delta_{th}\rho_{cr}*\Omega_M$).

In order to be consistent of the simulation parameters, we take
$\delta_{th}=200$, Hubble constant $H_0=72kms^{-1}Mpc^{-1}$ for a
flat universe, and $\Omega_M=0.216$ for our cosmological model.
With these standard parameters, $r_{vir}$ is defined to be:
\begin{equation}
r_{vir}=163h^{-1}kpc \left(\frac{\delta_{th}
\Omega_M}{43.2} \right)^{-1/3}
\left(\frac{M_{vir}}{10^{12}h^{-1}M_{\odot}}\right)^{1/3}
\end{equation}
Only two independent parameters, C and $M_{vir}$, need to be
specified to define the values of all other halo quantities.

The reconstruction of the LMC's orbital parameters depends
less sensitively on the Galactic mass distribution at
distances much smaller than $r_p$ where most of the
baryonic matter resides. Nevertheless, we include the
baryonic matter's contribution to the Galactic potential
so that our model is self consistent, ie it can reproduce
the observed circular velocity of the local standard of
rest.  The baryonic components include the mass of central
black hole, the Galactic nucleus, bulge and an exponential disk
with scale length $r_d$. We follow the equation outlined
in Klypin et al. (2002) such that
\begin{eqnarray}
M_b(r) &=& M_{BH}
       +0.025M_{b,vir}[1-exp(-2.64r^{1.15})]  \notag  \\
       &&+0.142M_{b,vir}[1-(1+r^{1.5})exp(-r^{1.5})]   \notag  \\
       &&+0.833M_{b,vir}\left[1-\left(1+\frac{r}{r_d}\right)
       exp\left(-\frac{r}{r_d}\right)\right]
\end{eqnarray}
The disk surface density is:
\begin{equation}
\Sigma(r)=\Sigma_0 \ exp \left(-\frac{r}{r_d} \right)
\end{equation}
where $M_{b,vir}$ is the total mass of the cooled baryons, as a fraction
of the virial mass of dark matter halo, $M_{b,vir}=0.05M_{vir}$
and $r_d=3.5kpc$ is the scale length of the disk corresponding
to the location of the Sun as $8.5kpc$. e.g. if $M_{vir}=10^{12}M_{\odot}$,
the total mass of the disk is $4.2 \times 10^{10}M_\odot$ and the bulge
mass as $7.1 \times 10^{9}M_\odot$.

Since the pericenter of the LMC's orbit is about $50kpc$, which is
far away from the typical size of disk and bulge, the exact baryonic mass
distribution is not very important for the
orbital history, but it is relevant to the solar circular velocity.
The rotation curve of an exponential disk is (Freeman 1970):
\begin{equation}
V^2_{disk}(r)=4{\pi}G\Sigma_0(y)[I_0(y)K_0(y)-I_1(y)K_1(y)]
\end{equation}
where $y=r/r_d$ and $I_n$,$K_n$ are the Modified Bessel
Functions of the first and second kind(Binney \& Tremaine 2008).
The zero-pressure rotation velocity can be determined as:
\begin{equation}
V^2_{c}(r)=V^2_{disk}(r)
+\frac{G}{r}[M_{BH}+M_{nul+bul}(r)+M_{halo}(r)]
\end{equation}
Here we do not consider the adiabatic contraction of the dark matter halos.

With this velocity of the Sun, we can calculate the current velocity of LMC.
If we choose $M_{vir}$ to be of the order $10^{12}M_\odot$ or $1.5 \times
10^{12}M_\odot$, it self-consistently gives $V_c$ of the Sun
to be $220$km s$^{-1}$ or $250$km s$^{-1}$ respectively from
the rotation curve (see Figs 3 and 4).
\begin{figure}
  \centering
  \includegraphics[width=0.9\linewidth,clip=true]{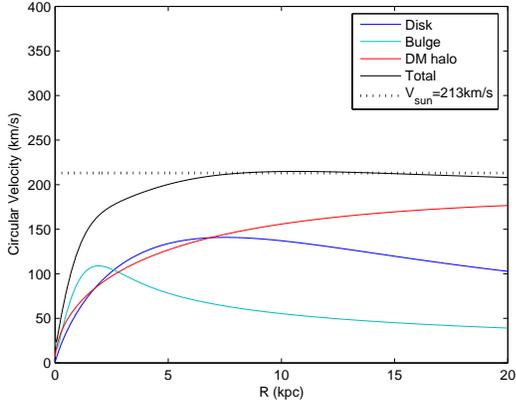}
\caption{The rotation curve for $M_{vir}=10^{12}M_\odot$,
$R_{vir}=200kpc$ and $C=12$. The total potential gives
$V_{c,sun}\sim213kms^{-1}$.}
  \label{fig:fig3}
\end{figure}
\begin{figure}
  \centering
  \includegraphics[width=0.9\linewidth,clip=true]{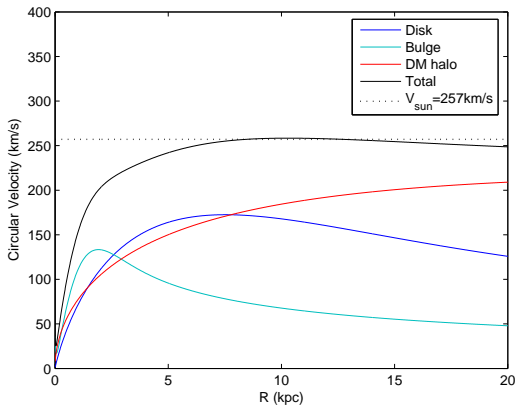}
\caption{The rotation curve for $M_{vir}=1.5 \times 10^{12}M_\odot$,
$R_{vir}=232kpc$ and $C=13$. The circular velocity at $8.5kpc$
is $V_{c,sun}\sim257kms^{-1}$.}
  \label{fig:fig4}
\end{figure}

When we choose different parameters of NFW models to calculate the orbital
history of LMC, the $V_c$ will also be adjusted slightly according to the
changing dark matter mass inside the location of the Sun.

\subsection{Simulated models of dark matter halo}

The NFW model is well suited for the part of the Galaxy within the
virial radius. If we extrapolate this model for the entire
dark matter halo, its density would decrease as $\rho{\propto}r^{-3}$
for $r > > r_s$. However, from the numerical simulations carried
out by Oser et al.(2010), we find that the actual halo density profile
is somewhat flatter than expected for the NFW model, ie there is
more residual dark matter at large distance than that inferred from the
NFW model. As we have discussed above, the LMC's orbital history is very
sensitively determined by the density gradient of the outside part of the
Galaxy. Here, we use a more realistic dark halo profiles which is extracted
directly from the cosmological simulations.

In order to constrain the dark matter density profiles
at large Galactocentric radii, we choose a set of 4
'zoom-in' cosmological simulations of individual halos
(Oser et al. 2010).  The initial condition of this simulation is
based on a flat WMAP3 cosmology (Spergel et al. 2007) with model
parameters: h=0.72, $\Omega_b = 0.044, \Omega_{m} = 0.216,
\Omega_{\Lambda} = 0.74$ and $\sigma_8=0.77$. The initial slope
of the power spectrum is $n_s = 0.95$.

At redshift zero we identify halos with the help of a
friends-of-friends finder and choose relaxed halos with no massive
companion close-by for re-simulation. We trace back in time all
particles closer than two times the virial radius to the halo center
in any given snapshot and replace those particles with multiple dark
matter particles of lower mass while adding the small scale density
fluctuations with the help of GRAFIC2 (Bertschinger 2001)
The dark matter particles outside the region of interest are merged
(depending on their distance to the re-simulated halo) to reduce the
particle count and the simulation time. This way we are able to
simulate structure formation in the cosmological context at high
resolution in a reasonable amount of time.  To evolve the high
resolution initial conditions from redshift $z=43$ to the present day
we use the parallel Tree-code GADGET-2 (Springel 2005).

The set of 4 halos in Table 1 have masses in the range
of $ 0.5 - 2 \times 10^{12} M_{\odot} h^{-1}$.  The computational
domain corresponds to a cosmic cube with the size of $100 \rm Mpc/h$.
The dark matter is represented by particles with a mass
$2.5 \times 10^7 M_{\odot} h^{-1}$ and a comoving gravitational
softening length of $890 h^{-1}$ pc. The radial dark halo density
structure is well resolved in the region from $r_p = 50$ kpc
to the virial radius and beyond.

In order to apply the simulation data to the computation of the LMC's
orbit, we first match the simulated dark matter distribution inside
200 kpc with NFW models to extract the best fitted values of $\rho_s$
and $r_s$.  Based on these two values, we then match the
density distribution at around 200 kpc with the fitting formula
\begin{equation}
\rho=\frac{\rho_s}{(r/r_s)(1+r/r_s)^\beta}
\end{equation}
to determine the value of $\beta$.  Instead of the idealized NFW profile in
equation (8) which is equivalent to the case of $\beta=2$,
we use the above fitting procedures to approximate the dark matter
halo's density profile beyond 200 kpc (see Table 1). In Figure 5, we
show the LMC's orbital history for various model parameters. Since the
$M_{vir}$ values for these halos are relatively high and $\beta <2$,
it is possible for LMC to have more than one peri-galacticon passage
even for $V_{sun}<250km/s$.  Note that there is one low-$M_{vir}$ model (1167), where the
LMC is undergoing its first peri galacticon passage.

\begin{figure}
  \centering
  \includegraphics[width=\linewidth,clip=true]{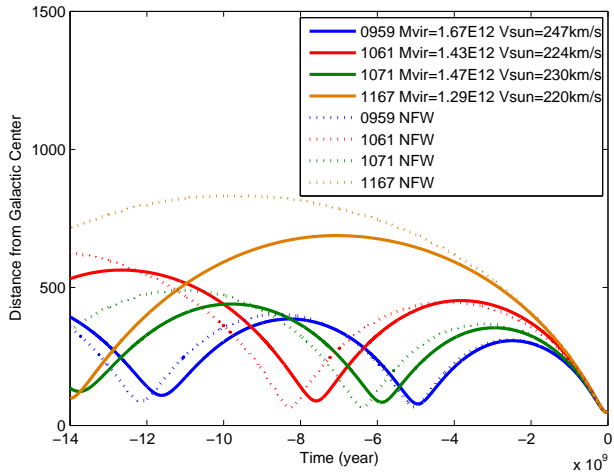}
\caption{Orbital history of the LMC for simulated dark matter halos.
The dark matter structure beyond $200kpc$ is determined from equation (18).
 The solid lines show the orbits for dark halos with $\beta$ as given in Table 1.
 The dotted curves show the situation if one would take a dark halo with
 similar virial parameters but fixing beta at 2 as expected for an idealized NFW profile.}
  \label{fig:fig5}
\end{figure}
\begin{deluxetable}{cccccccc}
    \tablecolumns{6}
    \tablehead{\colhead{data} & \colhead{$M_{vir}(10^{12}M_{\odot})$} & \colhead{$r_{vir}(kpc)$} & \colhead{$r_s(kpc)$} & \colhead{$\beta$} & \colhead{$V_c$
(km s$^{-1})$}}
    \startdata
          0959 & 1.67 & 240  & 21 &  1.78  & 247\\
          1061 & 1.43 & 228  & 18 &  1.61  & 224\\
          1071 & 1.47 & 231  & 21 &  1.69  & 230\\
          1167 & 1.29 & 221  & 21 &  1.83  & 220
    \enddata
    \tablenotetext{}{From the simulation results of Oser, et al. (2010).}
    \label{tab:simulation}
\end{deluxetable}

In the results listed in Table 1, we also calculate the theoretical
circular velocity of the Sun, in accordance with the Equations (16)
and (17). The actual observed circular velocity at the Sun's location
is related to the precise mass structure of the baryonic mass, because
for the inner part of the Galaxy, the contribution of circular velocity
from the baryonic matter in the disk is comparable to that from dark
matter (Fig 3 \& 4). The theoretical model we used here for an
exponential disk may lead to an underestimation of the Sun's circular
velocity. Although the theoretical circular velocities for the models
listed in Table 1 do not match exactly the latest observational
data (ie $V_c\sim254km/s$), the extra mass we need to make the velocity
to be consistent with the observation is very small compared with the
total mass within $50kpc$. Since the LMC does not venture into this inner
region of the Galaxy during its orbital history, the precise structure
inside $8.5kpc$ does not affect our determination of the LMC's orbital
history.

The observationally determined values of $V_c$ remain uncertain
within the range between 220 km s$^{-1}$ to 250 km s$^{-1}$ (Reid
et al. 2009a, Bovy et al. 2009, 2012, also see \S1). If these
values are applied to the NFW models, the lower limit of $V_c$ would imply
a single perigalacticon passage for the LMC whereas with the upper limit of
$V_c$, the LMC would have had multiple perigalacticon passages.  However,
with the modified density distribution for the outer galactic halo,
the LMC in three models (0959, 1061, and 1071) with $V_c > 228$ km s$^{-1}$
have had multiple perigalacticon passages.  Only in one model (1167) with
$V_c = 221$ km s$^{-1}$, the LMC's orbital period is marginally smaller
than 13 Gyr.  Based on these results, we suggest if $V_c > 228$ km
s$^{-1}$, LMC would likely have undergone more than one
perigalacticon passages.

The correction resulting from the most recent observed circular
velocity of the Sun leads to a smaller transverse velocity, orbital
period, and apo galacticon distance for the LMC (see \S2). However,
even without this change in the circular velocity of the Sun, the 1$\sigma$
error bar in the LMC's observed velocity is $\sim 18$km s$^{-1}$.
In order to assess the implication of these uncertainties on the
LMC's orbital history, we use a range of transverse velocities of the LMC's
current orbit that are consistent with the uncertainties in $V_t$ of$367\pm18$km s$^{-1}$ and
re-calculate the orbital history with the simulated halo based from
the model 1167 which has $M_{vir}~\sim 1.3 \times 10^{12}M_\odot$
and $R_{vir}\sim220$kpc.  In Figure 6, we show that the velocity
uncertainties are large enough to introduce a significant modification
in the LMC's orbital period as we have already suggested in \S3.2.
Although the mean and high values of the transverse velocity imply
that the LMC has just passed through its perigalacton for the first time,
the lowest value of the transverse velocity would imply a multiple peri
galacticon passage for the LMC.  This error analysis poses a
challenge to the statistical significance of the single passage
result obtained by Besla et al (2007).

\begin{figure}
  \centering
  \includegraphics[width=0.9\linewidth,clip=true]{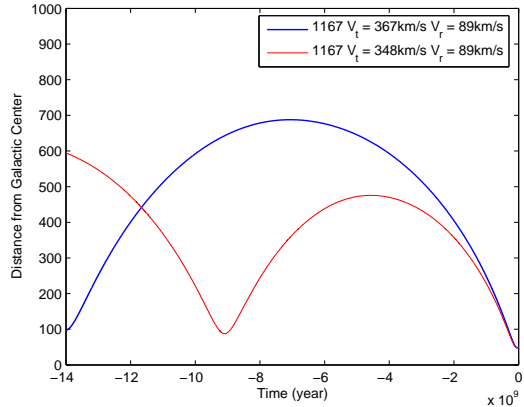}
\caption{Orbital history of the LMC with the minimum transverse velocity
$V_t = 348$ km s$^{-1}$.  This value of $V_t$ is within the error-bar
from that ($V_t=367$km s$^{-1}$) derived with $V_c=220$km s$^{-1}$.
For both cases, we used the same current radial velocity
for the LMC as $V_r=89$km s$^{-1}$ and the halo potential in
accordance to model 1167. }
  \label{fig:fig6}
\end{figure}

\subsection{Model including halo evolution}

Since the LMC's orbital period is either comparable (in the
case of a single passage) or a significant fraction (in the
case of multiple passages) of the Hubble time,
it is relevant to consider the evolution of the Galactic
dark matter halo. However, the build up of the dark matter
halo is a very complex process.  It involves merger events,
accretion and dynamical relaxation. In order to construct
a model for the evolution of the Galactic halo, we need to
identify the dominant mechanism.

We begin with an observational interpretation.  The age of
the thin Galactic disk can be traced by the age ($> 10$ Gyr) of
old white dwarfs in it(Hansen et al. 2004). The preservation of the thin disk
structure for this population of stellar remnants suggests that
the Galaxy may have not had any major merger events during the
past 10 Gyr. However, a significant mass increase due to a smooth
infall or cold streams cannot be ruled out.

Another approach to model the evolution of the dark matter halo
is to utilize the $\Lambda$CDM models which were used to model
the density distribution we used in \S4.1.  Here we adopt a
recently developed  prescription (Krumholz \& Dekel 2012) based
on the assumption that the in-streaming of baryonic and dark
matter has led to the growth in the Galactic potential and the
virial mass at a rate
\begin{equation}
\dot{M}_{vir,12}=-0.628M^{1.14}_{vir,12}\dot{\omega}
\end{equation}
where $M_{vir,12}=M_{vir}/10^{12}M_{\odot}$.  In the above equation,
$\omega=1.68/D(t)$ is the self-similar time variable of the
extended Press-Schechter (EPS) formalism, and $D(t)$ is the linear
fluctuations' growth function.  Based on this EPS formalism,
Neistein et al. (2006) and Neistein \& Dekel (2008) estimate
\begin{equation}
\dot{\omega}=-0.0476[1+z+0.093(1+z)^{-1.22}]^{2.5}Gyr^{-1}.
\end{equation}

The virial radius is related to both virial mass and redshift
via $r_{vir}{\propto}M_{vir}^{1/3}/H(z)^{2/3}$ (Burkert et al. 2010),
where $H(z)$ is:
\begin{equation}
H(z)=H_0[\Omega_{\Lambda}+(1-\Omega_{\Lambda}-\Omega_M)
(1+z)^2+\Omega_M(1+z)^3]^{\frac{1}{2}}.
\end{equation}
The redshift is also related to universal time through:
\begin{equation}
t=\frac{2}{3H_0{\Omega_0}^{1/2}(1+z)^{3/2}}.
\end{equation}

With these two relations, we can model the time evolution of
the virial radius and calculate the orbital history of the
LMC with this simplified evolution model for the dark matter halo.

We consider a low-mass (with the virial mass $M_{vir}
=1.3 \times 10^{12}M_{\odot}$) and a high-mass
($M_{vir}=1.66 \times 10^{12} M_{\odot}$) case.  These limits
are comparable with the minimum and maximum
masses of the simulation data.  For both cases, we set the
concentration parameter $C=12$ and neglect its evolution.
For these two sets of model parameters, the initial mass accretion
rate is relatively high ( $>100M_{\odot}$ yr$^{-1}$) at redshift $z>2$(Fig.7).
\begin{figure}
  \centering
  \includegraphics[width=0.9\linewidth,clip=true]{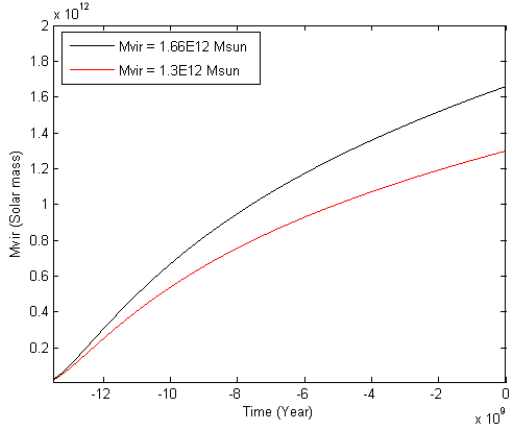}
\caption{Mass evolution according to eq.(19).}
\label{fig:fig7}
\end{figure}
Although in the evolutionary models, $M_{vir}$ increased steadily
LMC's orbits since the previous perigalacticon passage for the
evolutionary models are similar to those with a fixed halo.
For the low mass model (Fig. 8), the previous apo and perigalacticons
occurred at 4 and 8 Gyr ago when $M_{vir}$ has already attained
85\% and 50\% of its value at $z=0$ respectively.  The late mass increase of the
halo potential does not modify LMC's orbit significantly. However
more than 10Gyr ago, $M_{vir}$ was substantially smaller than today.
The play back integration of LMC's orbits for the evolving and
the non evolving halo models therefore diverges at high redshift.  This divergence
is similar to that between models which include or neglect contributions
from dynamical friction. However the dynamical friction effect
decreased at early epochs in the evolving halo model due to a
decreased density of dark matter back then.

Similar results are also obtained for the high mass model (Fig. 9).
In this case, the previous apo and perigalacticons
occurred at 1.8 and 3.6 Gyr ago when $M_{vir}$ has already attained
94\% and 88\% of its value at $z=0$.  Thus, the divergence between
the orbits for the evolving and non evolving dark matter halo is only
significant more than 8 Gyr ago, when the LMC passed through two peri
galacticon passages prior to the present epoch.  We note that
when the evolution of the halo is taken into account, the number
of passages declines by at most one in comparison with that
for the non evolving halo.

\begin{figure}
  \centering
  \includegraphics[width=0.9\linewidth,clip=true]{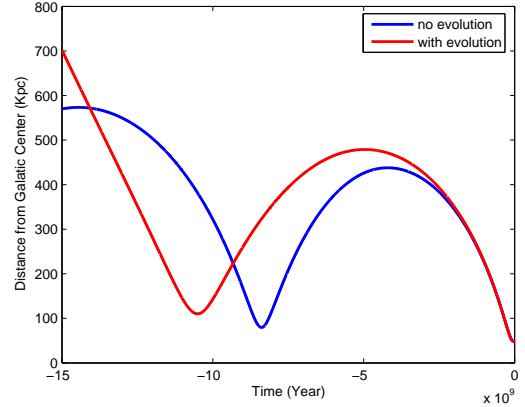}
\caption{Orbital history of the LMC with and without evolution of
the dark matter halo.  The concentration parameter is $C=12$
as constant with the current $M_{vir}=1.3 \times 10^{12}M_{\odot}$
and kept constant at a value of $R_{vir}=221kpc$.
The corresponding circular velocity is $V_{sun}=236$km s$^{-1}$.}
\label{fig:fig8}
\end{figure}

\begin{figure}
  \centering
  \includegraphics[width=0.9\linewidth,clip=true]{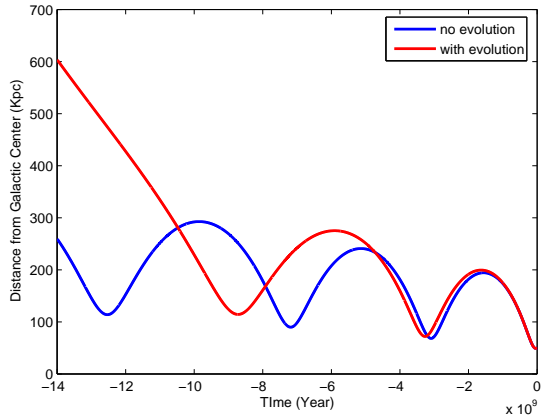}
\caption{Orbital history of the LMC with and without the evolution
of a larger dark matter halo.
The concentration parameter is $C=12$ and constant with the current
$M_{vir}=1.66*10^{12}M_{\odot}$ and $R_{vir}=240kpc$.
The corresponding circular velocity is $V_{sun}=259$km s$^{-1}$.}
\label{fig:fig9}
\end{figure}

In the above models, we adopt a constant concentration parameter.
For an alternative possibility, we consider a model in which
$C$ increased linearly from $C=4$ to $C=12$ during cosmic time
for the evolving halo models.  We consider the case with the
asymptotic virial mass $M_{vir}=1.66*10^{10}M_{\odot}$ (see Fig.10).
The small concentration at early time makes the LMC's orbit more
loosely bound. However, in comparison with the non-evolving halo
model (with a fixed $C=12$), the difference results from $C$ variations
is not very significant. It seems that the orbital period of LMC is more
dependent on the current concentration of the dark matter halo than the
initial value.

\begin{figure}
  \centering
  \includegraphics[width=0.9\linewidth,clip=true]{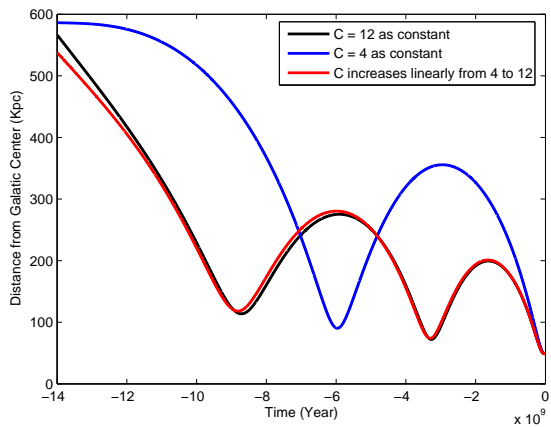}
\caption{Orbital history of LMC with linear evolution of
concentration parameter of DM halo, compared to the results
with fixed $C=12$ and $C=4$. The final mass is $M_{vir}=1.66
\times 10^{12}M_{\odot}$.}
\label{fig:fig10}
\end{figure}

\section{Summary}

Besla et al (2007) utilized LMC's three dimensional velocity data
to reconstruct LMC's orbital history.  They made a bold suggestion
that the LMC has passed the perigalacticon of its Galactic orbit
for the first time.  This conclusion, if it can be verified, would
invalidate the tidal disruption hypothesis for the Magellanic Stream.

We show that these previous results depend critically on the LMC's
transverse velocity and the dark matter distribution in the outer
Galactic halo.  At the LMC's latitude and longitude in the sky,
a large fraction of its observed proper motion is due to the Sun's
circular velocity around the Galactic center.  We use the most
recently measured solar circular velocity (254 km s$^{-1}$) rather
than its conventional value (220 km s$^{-1}$) to deduce a smaller
transverse speed, apogalacticon distance, and orbital period for
the LMC.  It raises the possibility that the LMC actually had two
or more previous perigalacticon passages.

We adopt the current velocity of LMC with a different solar velocity,
and a simply modeled potential profile of the Milky Way outside 50kpc
to examine its orbital history. We find that the number of passages
depends sensitively on the slope of the potential and density
distribution of the dark matter halo as well as the current transverse
velocity of LMC near its perigalacticon distance.

With an illustrative model (in Fig. 2), we show that, for a
relatively flat Galactic potential (with lambda=0.1), a modest
(7 percent) decrease in the deduced value of $V_t$ can
lead to a much larger (30 percent) decline in the LMC's inferred
orbital period.  Thus, a precise determination of LMC's orbital
history can provide a sensitive measurement on the distribution
of dark matter halo in the extended outer regions of the Milky
Way Galaxy.

In this paper, we have neglected the perturbation on the LMC from any
other member of Local Group, including M31. If we choose a massive
model of dark matter halo ($M_{vir}=1.66 \times 10^{12}M_{\odot}$), the
pericenter's location of the last two orbital periods would be well
within $400kpc$ (Fig.9). The current distance between M31 and the Milky
Way is about $700kpc$, it was even further away in the past since they
are currently approaching each other. The M31-Galaxy orbital plane
also appears to be inclined to the LMC-Galaxy orbital plane.
It's perturbation on the previous the LMC's orbital history
appear to be weak. This assumption is also consistence from the
lack of warp along the great circle which is traced out by the
Magellanic Stream.  Nevertheless, a more comprehensive set of simulations
is needed to justify the dynamic independence between M31 and the Milky
Way galaxies.

\section{Acknowledgement}
We thank Jim Peebles, Scott Tremaine, and an anonymous referee for useful
comments and suggestions. And this work is supported
in part by NASA grant NNX08AL41G.

\section{Reference}
\scriptsize{
Besla, G., Kallivayalil, N., Hernquist, L., Roberston, B., Cox, T.
J., van der Marel, R. P., \& Alcock, C. 2007, ApJ, 668, 949

Besla, G., Kallivayalil, N., Hernquist, L., van der Marel, R. P.,
Cox, T.J., \& Keres, D. 2010, ApJL, 721, L97

Bertschinger, E. 2001, ApJS, 137, 1

Binney, J., Tremaine, S. 1987, Galactic Dynamic, (Princeton, NJ;
Princeton University Press)

Binney, J., \& Tremaine, S. 2008, in Galactic Dynamics, Second Edition
(Princeton, NJ: Princeton Univ. Press), 390

Bovy, Jo, Hogg, David W., Rix, Hans-Walter 2009, ApJ, 704, 1704

Bovy, Jo, Prieto, C.A., Beers, T.C., Bizyaev, D., da Costa, L.N.,
Cunha, K., Ebelke, G.L., Eisenstein, D.J., Frinchaboy, P.M.,
Ṕere, A.L.G.,  Girardi, L., Hearty, F.R., Hogg, D.W., Holtzman, J.,
Maia, M.A.G., Majewski, S.R., Malanushenko, E., Malanushenko, V. ,
M ́esza ́r, S., Nidever, D.L., O’Connell, R.W., O’Donnell, C.,
Oravetz, A., Pan, K., Rocha-Pinto, H.J., Schiavon, R.P., Schneider,
D.P., Schultheis, M., Skrutskie, M., Smith, V.V., Weinberg, D.H.,
Wilson, J.C., \& Zasowski, G. 2012, ArXiv:1209.0759

Burkert, A., Genzel, R., Bouch, N., Cresci, G., Khochfar, S.,
Sommer-Larsen, J., Sternberg, A., Naab, T., Forster Schreiber, N.,
Tacconi, L., et al. 2010, ApJ, 725, 2324

Chandrasekhar, S. 1943, ApJ, 97, 255

Dehnen,W., \& Binney, J. J. 1998, MNRAS, 298, 387

Fellhauer, M., Lin, D. N. C. 2007, MNRAS, 375, 604

Freeman, K. C. 1970, ApJ, 160, 811

Fujimoto, M., Sofue, Y. 1976, A$\&$A, 47, 263

Gardiner, L. T., Sawa, T., Fujimoto, M. 1994, MNRAS, 266, 567

Gunn, J. E., Gott, J. Richard, III 1972, ApJ, 176, 1

Hansen, Brad M. S., Richer, Harvey B., Fahlman, Greg. G., Stetson, Peter B., Brewer, James, Currie, Thayne, Gibson, Brad K., Ibata, Rodrigo, Rich, R. Michael, Shara, Michael M. 2004, ApJS, 155, 551

Hashimoto, Y., Funato, Y., \& Makino, J. 2003, ApJ, 582, 196

Jones, B. F., Klemola, A. R., Lin, D. N. C. 1994, AJ, 107, 1333

Kahn, F. D., Woltjer, L. 1959, ApJ, 130, 705

Kallivayalil, N., van der Marel, R. P., Alcock, C.,
Axelrod, T., Cook, K. H., Drake, A. J., Geha, M. 2006, ApJ, 638, 772

Kallivayalil, N., van der Marel, R. P., Alcock, C. 2006, ApJ, 652, 1213

Klypin, A., Zhao, H., \& Somerville, R. S. 2002, ApJ, 573, 597

Krumholz, M. R. \& Dekel, A. 2012, ApJ, 753, 16

Lin, D. N. C., Jones, B. F., Klemola, A. R. 1995, ApJ, 439, 652

Lin, D. N. C., Lynden-Bell, D. 1982, MNRAS, 198, 707

Lynden-Bell, D., Lin, D. N. C. 1977, MNRAS, 181, 37

Majewski, S. R., Skrustkie, M. F., Weinberg, M. D., \& Ostheimer, J. C. 2003, ApJ, 599, 1082

Mastropietro, C., Moore, B.; Mayer, L., Stadel, J. 2005, ASPC, 331, 89

Mathewson, D. S., Cleary, M. N., \& Murray, J. D. 1974, ApJ, 190, 291

Meatheringham S.J., Dopita M.A., Ford H.C. \& Webster B.L. 1988, ApJ, 327,651

Mirabel, I. F., Turner, K. C. 1973, A$\&$A, 22, 437

Moore, B., Davis, M. 1994, MNRAS, 270, 209

Murai, T., Fujimoto, M. 1980, PASJ, 32, 581

Navarro, J. F., Frenk, C. S., White, S. D. M. 1997, ApJ, 490, 493

Neistein, E., \& Dekel, A. 2008, MNRAS, 383, 615

Neistein, E., van den Bosch, F. C., \& Dekel, A. 2006, MNRAS, 372, 933

Oser, L., Ostriker, J.P., Naab, T., Johansson, P.H. \& Burkert, A. 2010, ApJ, 725, 2312

Peebles, P. J. E. 2001, ASPC, 252, 201

Peebles, P. J. E. 2010, arXiv:1009.0496

Putman, M. E., Gibson, B. K., Staveley-Smith, L. 1999, ASPC, 165, 113

Reid, M. J., Menten, K. M., Zheng, X. W., Brunthaler, A., Moscadelli, L.,
Xu, Y., Zhang, B., Sato, M., Honma, M., Hirota, T. et al. 2009a, ApJ, 700, 137

Shattow, G. \& Loeb, A. 2009, MNRAS, 392, L21.

Spergel, D. N., Bean, R., Dor¨¦, O., Nolta, M. R., Bennett, C. L., Dunkley, J.,
Hinshaw, G., Jarosik, N., Komatsu, E., Page, L.,  et al. 2007, ApJS, 170, 377

Springel, V., Yoshida, N., \& White, S. D. M. 2001, New Ast., 6, 79

Springel, V., Yoshida, N., \& White, S. D. M. 2005, MNRAS, 364, 1105

Toomre, A., 1972, QJRAS, 13, 266

van der Marel, R. P., Alves, D. R., Hardy, E., Suntzeff, N. B.
2002, AJ, 124, 2639

van der Marel, R. P., Fardal, M., Besla, G., Beaton, R. L.,
Sohn, S. T., Anderson, J., Brown, T., Guhathakurta, P. 2012, ApJ, 753, 8
}

\end{document}